%Paper: gr-qc/9210014
%From: rhb@het.brown.edu (Robert Brandenberger)
%Date: Sat, 24 Oct 92 15:41:01 EDT

\input phyzzx
\hsize=6.5in
\hoffset=0.0in
\voffset=0.0in
\vsize=8.9in
\FRONTPAGE
\line{\hfill Brown-HET-876}
\line{\hfill September 1992}
\vskip1.5truein
\titlestyle{{ A NONSINGULAR UNIVERSE}\foot{Invited talk at International
School of Astrophysics ``D.
Chalonge'', 2nd course, 6-13 September 1992, Erice, Italy, to be published in
the proceedings (World Scientific, Singapore, 1992).}\break}
\bigskip
\author{Robert H. Brandenberger}
\centerline{{\it Physics Department}}
\centerline{{\it Brown University}}
\centerline{{\it Providence, RI  02912, USA}}
\bigskip
\centerline{ABSTRACT}
We construct an effective action for gravity in which all homogeneous solutions
are nonsingular$^{1)}$.  In particular, there is neither a big bang nor a big
crunch.  The action is a higher derivative modification of Einstein's theory
constructed in analogy to how the action for point particle motion of particles
in special relativity is obtained from Newtonian mechanics.
\vfill\endpage
\noindent \undertext{1.~~Introduction}

The singularity theorems of general relativity$^{2)}$ prove that space--time
manifolds in general relativity are geodesically incomplete.  The key
assumptions which are used in the proofs of these theorems are that the action
of gravity is an unmodified Einstein action, and that mattter satisfies the
energy dominance condition $\epsilon > 0$ and $\epsilon + 3p \geq 0$, where
$\epsilon$ and $p$ are energy density and pressure respectively.

Whereas the singularity theorems do not provide any general information about
the nature of the singularity, well known examples of space--time show that at
 a singularity typically some of the curvature invariants $R\, ,\> R_{\mu
\nu}\>R^{\mu \nu}\, ,\>R_{\alpha\, \beta\, \gamma \, \delta}\> R^{\alpha\,
\beta\, \gamma\, \delta}\, , \dots$ blow up.  These singularities should be
viewed as a breakdown of general relativity at high curvatures.

Quantum gravity and alternative fundamental theories of all four forces, such
as string theory, have been invoked as ways towards a solution of the
singularity problems.  As yet these avenues have not been developed far enough
to provide a solution. We believe, however, that a new fundamental theory which
includes gravity will solve the singularity problem. One way
towards realizing a singularity free theory is to guess an effective action for
gravity which contains no singularities. Constructing such a theory is what
we attempt in this work.

As a preliminary, recall that the well known and successful theories of special
relativity (SR) and quantum mechanics are based on inequalities $v < c$ and
$\Delta x\Delta p \geq \hbar$ respectively.  Hence, the hope arises that it
might be possible to construct a new theory of gravity based on an inequality
involving Newton's constant $G$.  For example, in a theory with a fundamental
length $\ell_f$ (i.e. $\ell \geq \ell_f$ for all lengths $\ell$), all curvature
invariants are automatically bounded ($R\leq \ell_f^{-2}\, ,\>R_{\mu\nu}\,
R^{\mu\nu}\, \leq\, \ell_f^{-4}\, ,$ etc.).  However, a fundamental length is
incompatible with a continuum theory of space and time, and thus we will
attempt to realize the constraints on the curvature invariants directly..

Our goal is to construct a theory in which \undertext{all} curvature invariants
are bounded and in which space--time is geodesically complete.  This formidable
problem can be reduced substantially by invoking the  ``Limiting Curvature
Hypothesis''$^{3)}$, according to which one
\item{i)} finds a theory  in which a small number of invariants is explicitly
bounded, and
\item{ii)} when these invariants take on their limiting values, a definite
nonsingular solution (namely de Sitter) is taken on.

\noindent As a consequence of the limiting curvature hypothesis, automatically
all invariants are bounded, and space--time is geodesically complete in its
asymptotic regions.

The limiting curvature hypothesis has interesting consequences for Friedmann
models and for spherically symmetric vacuum space--times$^{4)}$.  A collapsing
Universe will not reach a big crunch, but will end up as a contracting de
Sitter Universe $(k=0)$ or a de Sitter bounce $(k=1)$ followed by re-expansion
(see Fig. 1). For a spherically symmetric vacuum solution, there would be no
singularity inside the Schwarzschild horizon; instead, a de Sitter Universe
will be reached when falling through the horizon towards large curvature (see
Fig. 1.).
\midinsert
\vskip 12cm
\hsize=6in \raggedbottom
\noindent{\bf Fig. 1:} Penrose diagrams for a collapsing Universe (left) and
for a black hole (right) in Einstein's theory and after implementing the
limiting curvature hypothesis (bottom). Wavy lines denote a singularity (in the
case of the collapsing Universe the big crunch), the symbols C, DS and E stand
for collapsing phase, de Sitter phase and expanding phase respectively, and H
denotes the Schwarzschild horizon.
\endinsert
\endpage
\noindent \undertext{2.~~ Construction}

In order to realize the limiting curvature hypothesis, we must abandon at least
one of the assumptions of the Penrose--Hawking theorems.  Unlike in
inflationary cosmology$^{5)}$ we do not invoke ``strange'' matter which
violates the energy dominance condition.  Instead, we drop the assumption that
gravity is described by a pure Einstein action.

The theory discussed here is a higher derivative modification of Einstein
gravity.  It is reasonable to consider such modifications since the Einstein
theory is known to break down at high curvatures -- based on perturbative
quantum gravity calculations, quantum field theory effects in curved
space--time, and on taking low energy limits of fundamental theories of all
forces such as string theory.

Most higher derivative gravity models have much worse singularity properties
than the Einstein theory. Hence, it is a nontrivial task to construct
\undertext{a} model which has better properties.  As an added bonus, the
construction which leads to our nonsingular Universe is well motivated in
analogy to how the action for particle motion in special relativity emerges
from the point particle action in Newtonian mechanics.

Special relativity is a theory in which point particle velocities $v$ are
bounded.  Starting from Newtonian mechanics (in which $v$ is unbounded) for
which the point particle action is
$$
S=m \int dt\>  {1\over 2}\>{\dot x}^2\, ,\eqno(2.1)
$$
$m$ being the particle mass, the action for special relativity can be
obtained$^{6)}$ using a Lagrange multiplier construction
$$
S = m \int dt \>\left [{1\over 2} \>{\dot x}^2 + \varphi\>{\dot x}^2 -
V(\varphi)\right ]\, .\eqno(2.2)
$$
Provided that $V(\varphi)$ increases no faster than $\varphi$ at large
$\varphi$, the quantity which couples to $\varphi$, namely ${\dot x}^2$, is
automatically bounded, as follows from the variational equation with respect to
$\varphi$
$$
{\dot x}^2 = {\partial V\over {\partial \varphi}}\, .\eqno(2.3)
$$
In order to recover the correct Newtonian limit at low velocities, $V(\varphi)$
must be proportional to $\varphi^2$ as $\varphi \rightarrow 0$.  Thus, the
conditions on $V(\varphi)$ are
$$
V(\varphi) \sim \cases{\varphi &$\vert \varphi \vert \rightarrow \infty$\cr
\varphi^2 &$\varphi \rightarrow 0\, .$\cr}\eqno(2.4)
$$
Up to factors of 2, the simplest potential which satisfies (2.4) is
$$
V(\varphi) = {2 \varphi^2\over {1 +2\varphi}}\eqno(2.5)
$$
Eliminating the Lagrange multiplier $\varphi$ via (2.3) and substituting into
(2.2), the action of a point particle in special relativity
$$
S = \int dt \>\sqrt{1 - {\dot x}^2}\eqno(2.6)
$$
results.

Our idea$^{1)}$ is to imitate the above construction in gravity.  Starting with
Einstein's theory of general relativity with action
$$
S = \int d^4 x\>\sqrt{-g}\>R\eqno(2.7)
$$
and unbounded Ricci scalar curvature, we construct a new gravity theory by
introducing a Lagrange multiplier $\varphi_1$, with potential $V_1 (\varphi_1)$
which couples to $R$, the quantity we wish to bound:
$$
S = \int d^4 x\>\sqrt{-g}\>\left [R+\varphi_1 R + V_1 (\varphi_1)\right]\,
.\eqno(2.8)
$$
The potential $V_1 (\varphi_1)$ must satify the same asymptotic properties as
given in (2.4).

However, the action (2.8) is not sufficient.  In order to obtain a nonsingular
Universe, we must implement the limiting curvature hypothesis.  This is
achieved once again by using the Lagrange multiplier technique.  At this point
we restrict our attention for the moment to homogeneous and isotropic
space--times.

Consider the invariant
$$
I_2 = 4R_{\mu\nu}\>R^{\mu\nu} - R^2\, .\eqno(2.9)
$$
This invariant is positive semidefinite, and vanishes only if space--time is de
Sitter.  Hence, we will implement the limiting curvature hypothesis by forcing
$I_2$ to zero at high curvatures.  We chose
the action
$$
S = \int d^4 x \>\sqrt{-g}\>\left [R +\varphi_1 R +\varphi_2 \sqrt{I_2} + V_1
(\varphi_1) + V_2 (\varphi_2)\right ]\, .\eqno(2.10)
$$
Provided that
$$
V_2 (\varphi_2) \sim \cases{{\rm const} &$\vert \varphi_2 \vert\rightarrow
\infty$\cr
\varphi_2^2 & $\varphi_2 \rightarrow 0$\cr}\eqno(2.11)
$$
then for $\vert \varphi_2 \vert \rightarrow \infty$ space--time becomes de
Sitter, and the low curvature limit of the theory agrees with general
relativity.

By construction, a theory with action (2.10) becomes de Sitter at large
$\varphi_2$.  It remains to be shown that there are no singularities for finite
values in the $\varphi_1 / \varphi_2$ phase space.  To show this, we need a
specific model.
\bigskip
\noindent \undertext{3.~~Specific Model}

As the most simple realization of a nonsingular Universe we consider the
action$^{1)}$
$$
S = \int d^4 x\>\sqrt{-g}\>\left [ (1 +\varphi_1) R -\left(\varphi_2 + {6\over
{\sqrt{12}}}\>\varphi_1\right )\, I_2^{1/2} + V_1 (\varphi_1) + V_2 (\varphi_2)
\right ]\eqno(3.1)
$$
with
$$
\eqalign{V_1(\varphi_1) &= 12 H_0^2\> {\varphi_1^2\over {1 +\varphi_1}}\>
\left( 1 - {\ell n (1+\varphi_1)\over {1+\varphi_1}}\right )\cr
V_2 (\varphi_2) &= -\, \sqrt{12}\> H_0^2\> {\varphi_2^2\over
{1+\varphi_2^2}}}\eqno(3.2)
$$
Apart from the logarithmic term in $V_1$, the above potentials are the most
simple ones which satisfy the asymptotic conditions (2.4) and (2.11).  It was
necessary$^{1)}$ to add the next leading (logarithmic) term in $V_1$ in order
to prevent trajectories from reaching $\varphi_1 \rightarrow \infty$ for $\vert
\varphi_2 \vert < 1$.

The general variational equations which follow from (3.1) are rather
complicated (see Ref. 7).  However, when applied to a collapsing Universe with
metric
$$
ds^2 = -\, dt^2 + a^2 (t) d{\bf{x}}^2 \eqno(3.3)
$$
and Hubble parameter
$$
H = {{\dot a}\over a} < 0\, ,\eqno(3.4)
$$
the variational equations become simple$^{1)}$:
$$
H^2 = {1\over {12}}\>V_1^\prime\, ,\eqno(3.5)
$$
$$
{\dot H} = -\>{1\over {\sqrt{12}}}\>V_2^\prime\eqno(3.6)
$$
and
$$
3 (1 - 2 \varphi_1)\, H^2 + {1\over 2}\, (V_1 + V_2) = {6\over
{\sqrt{12}}}\>H({\dot \varphi}_2 + 3 H\varphi_2)\, .\eqno(3.7)
$$
{}From (3.5) it follows that $\varphi_1 > 0$, from (3.6) that $\vert
\varphi_2\vert \rightarrow \infty$ is equivalent to de Sitter space, and (3.7)
can be combined with the time derivative of (3.5) and with (3.6) to yield
$$
{d\varphi_2\over {d\varphi_1}} = {V_1^{\prime\prime}\over {V_1^\prime\,
V_2^\prime}} \>\left [ -\, {1\over 4} (1 - 2\varphi_1)\, V_1^\prime + {1\over
2}\, (V_1 + V_2) + {3\over {2 \sqrt{12}}} \>V_1^\prime\>\varphi_2\right ]\,
,\eqno(3.8)$$
an equation from which the trajectories of this dynamical system in $\varphi_1
/ \varphi_2$ phase space can be read off.

The system of equations (3.5, 3.6 \& 3.8) must be analyzed to show that there
are no singular solutions.  The asymptotic regions $\vert \varphi_1 \vert\,
,\>\vert \varphi_2\vert \ll 1$ and $\vert \varphi_1 \vert\, ,\>\vert \varphi_2
\vert \gg 1$ can be analyzed analytically$^{1)}$.  It can be seen that there
are two types of solutions:  periodic solutions about Minkowski space
$(\varphi_1 = \varphi_2 = 0)$ and solutions which start and end at $\vert
\varphi_2\vert = \infty$, i.e. in de Sitter space (see Fig. 2).  It can be
shown numerically$^{7)}$ that there are indeed no singular points for finite
values of phase space and that the trajectories connect in a way which can be
guessed from the analytical analysis of the asymptotic regions.  Thus, we have
demonstrated that all solutions are nonsingular.

So far, only vacuum solutions of our new gravitational theory have been
discussed.  It is easy to include matter in the analysis by considering the
action
$$
S_{\rm full} = S + S_m\, ,\eqno(3.9)
$$
where $S$ is the gravitational action of (3.1), and $S_m$ is the action for
matter in the presence of the metric $g_{\mu\nu}$.  We have investigated$^{7)}$
the model obtained by adding hydrodynamical matter with an equation of state $p
= w\rho$ and $w=0$ (cold matter) or $w = 1/3$ (radiation).

The interesting result of our analysis$^{7)}$ is that for $\vert \varphi_2
\vert \rightarrow \infty$ the trajectories are unchanged when adding matter,
even though for a contracting spatially flat Universe the energy density is
increasing exponentially.

The only change for a spatially closed Universe is that the contracting de
Sitter phase is replaced by a de Sitter bounce.

In conclusion, we have presented a model in which all homogeneous and isotropic
solutions are nonsingular.
\endpage
\topinsert
\vskip 21cm
\hsize=6in \raggedbottom
\noindent{\bf Fig. 2:} Phase diagram for the solutions of (3.8), arrows
pointing in the direction of increasing time. As can be shown using (3.6), all
asymptotic solutions are de Sitter.
\endinsert

%\bigskip
%\vfill

%\endpage
\noindent \undertext{4. ~~Wild Speculations}

Since matter does not change the evolution of space--time at large curvatures,
the gravitational interactions are asymptotically free, i.e. the effective
coupling $G_{\rm eff}$ of matter to gravity tends to zero.  This is a first
very nice property of our model.

Secondly, when applied to an expanding Universe, our theory implies that it has
emerged from an initial de Sitter phase.  Thus, an inflationary period is
obtained without assuming the presence of matter violating the energy dominance
condition.  This result, however, is no surprise, since it is well known$^{8)}$
that higher derivative gravity models lead to inflation.

Let us now combine the first two results and consider the quantum generation of
density perturbations in the initial de Sitter phase.  These perturbations are
streched by inflation and may become the seeds for structures in the Universe.
In scalar field driven inflationary models, the magnitude of the scalar metric
fluctuations is too large without requiring that a particle physics parameter
(coupling constant of a $\lambda \varphi^4$ interaction term or a mass scale
$m$ in a theory of chaotic inflation with potential ${1\over 2} m^2\,
\varphi^2$) be artificially small.  However, since the magnitude of these
perturbations is proportional to $G_{\rm eff}$, it is conceivable that in our
model there will be no fine tuning for inflation.

Next, let us consider an application to black holes.  For black holes in
Einstein's theory of general relativity, Hawking radiation leads to its
evaporation with ever increasing speed.  However, the strength of Hawking
radiation is proportional to $G_{\rm eff}$.  Hence, in our theory Hawking
radiation may automatically shut off as the black hole mass decreases towards
its critical value $M_{\rm crit}$, which is in turn determined by when
curvature invariants like $C^2$ reach their limiting values $(H_0^4)$.

A consequence of the above is that black hole remnants will remain.  Hence,
there will be no loss of quantum coherence  in the presence of black holes
(when calculated in the semiclassical approximation).  Neither will there be
global charge violation by black holes.
\bigskip
\noindent \undertext{5.~~Extension to an Anisotropic Universe}

Hopefully the reader is at this point persuaded that it is worth while to
explore our theory further and see if the wild speculations mentioned in the
previous section can indeed be realized.

As a first step, we have explored$^{9)}$ whether our theory can damp out
anisotropy at high curvatures, such that asymptotically also an anisotropic
Universe will lead to de Sitter space.

Obviously, the action (3.1) with invariant $I_2$ given by (2.9) is
insufficient, since $I_2$ does not depend on the anisotropy.  However, we can
easily improve the prospects by changing $I_2$ to
$$
I_2 = 4 R_{\mu\nu}\>R^{\mu\nu} - R^2 + C^2\eqno(5.1)
$$
where $C^2 = C_{\mu\nu \rho\sigma }\> C^{\mu\nu \rho\sigma}$ and
$C_{\mu\nu\rho\sigma}$ is the Weyl tensor.  We maintain the form of the action
(3.1).

Based on our previous investigations, we should expect to be able to achieve
our goal.  As $\varphi_2 \rightarrow \infty$, the invariant $I_2$ is again
driven to zero.  This will imply (in cases when $C^2 \geq 0$ ) both
$$
C^2 = 0\eqno(5.2)
$$
and
$$
4 R_{\mu\nu}\>R^{\mu\nu} - R^2 = 0\, .\eqno(5.3)
$$
The condition (5.2) implies decrease in anisotropy, and then (5.3) tells us
that the asymptotic solution (which is homogeneous) will be de Sitter space.

To verify the above claims, we have considered$^{9)}$ the simplest anisotropic
Universe with metric
$$
g_{\mu\nu} = \pmatrix{-1 & & &\cr
 &a^2 e^{\beta (t)}& &\cr
 & &a^2 e^{\beta (t)} &\cr
 & & &a^2 e^{- 2 \beta (t)}\cr}\, .\eqno(5.4)
$$
The variational equations can be derived using a convenient trick:  we replace
the time--time component $g_{00}$ by $- \alpha (t)^2$, insert the metric into
(3.1) and vary with respect to $\alpha (t)\, ,\, a (t)\, ,\, \beta(t)\, ,\,
\varphi_1 (t)$ and $\varphi_2 (t)$.  Still, the resulting equations are rather
complicated.

It must be shown that for $\vert \varphi_2 \vert \rightarrow \infty$ the
anisotropy tends to zero, i.e. ${\dot \beta}\rightarrow 0$.  This can be done
by picking out the terms which dominate in the equations of motion in the limit
$\vert \varphi_2 \vert\rightarrow \infty$.  As demonstrated in Ref. 9, this is
indeed the case.
\endpage
\noindent \undertext{Conclusions}

We have presented an effective action for gravity based on a higher derivative
modification of Einstein's theory of general relativity in which all
homogeneous solutions are nonsingular.  All corresponding space--time manifolds
are geodesically complete and either approach de Sitter space asymptotically or
oscillate about Minkowski space.  We have speculated that in our theory also
singularities inside the black hole horizon might be avoided.

\noindent \undertext{Acknowledgements}

The results and ideas presented in this article are based on key ideas by and
joint work with V. Mukhanov.  I am also grateful to my collaborators M.
Mohazzab, A. Sornborger and M. Trodden.  I wish to thank N. Sanchez for
organizing a stimulating school and for inviting me to present this lecture.
\bigskip
\REF\one{V. Mukhanov, and R. Brandenberger, {\it Phys. Rev. Lett.} \undertext
{68}, 1969 (1992).}
\REF\two{R. Penrose, {\it Phys. Rev. Lett.} \undertext{14}, 57
(1965)\hfill\break
S. Hawking, Proc. {\it R. Soc. London} {\bf A} \undertext{300}, 182 (1967).}
\REF\three{M. Markov, {\it Pis'ma Zh. Eksp. Teor. Fiz.} \undertext{36}, 214
(1982); \undertext{46}, 342 (1987);\hfill\break
 V. Ginsburg, V. Mukhanov and V. Frolov, {\it Zh. Eksp. Teor. Fiz.}
\undertext{94}, 1 (1988).}
\REF\four{V. Frolov, M. Markov and V. Mukhanov, {\it Phys. Lett.} {\bf
B}\undertext{216}, 272 (1989); {\it Phys. Rev.} {\bf D}\undertext{41}, 383
(1990).}
\REF\five{A. Guth, {\it Phys. Rev.} {\bf D}\undertext{23}, 347 (1981).}
\REF\six{B. Altshuler, {\it Class. Quant. Grav.} \undertext{7}, 189 (1990).}
\REF\seven{R. Brandenberger, V. Mukhanov and A. Sornborger, in preparation
(1992).}
\REF\eight{A. Starobinskii, {\it Phys. Lett.} {\bf B}\undertext{91}, 99
(1980).}
\REF\nine{R. Brandenberger, M. Mohazzab, V. Mukhanov, A. Sornborger and M.
Trodden, in preparation (1992).}
\refout

\vfill\end